# Kondo effect in quantum dots coupled to ferromagnetic leads with noncollinear magnetizations: effects due to electron-phonon coupling

R. Świrkowicz[1], M. Wilczyński[1], J. Barnaś[2,3]

[1]Faculty of Physics, Warsaw University of Technology, ul. Koszykowa 75,

00-662 Warsaw, Poland

[2]Department of Physics, Adam Mickiewicz University, ul. Umultowska 85,

61-614 Poznań, Poland

[3]Institute of Molecular Physics, Polish Academy of Sciences, ul. Smoluchowskiego 17,

60-179 Poznań, Poland

**Abstract**

Spin-polarized transport through a quantum dot strongly coupled to ferromagnetic electrodes with non-collinear magnetic moments is analyzed theoretically in terms of the non-equilibrium Green function formalism. Electrons in the dot are assumed to be coupled to a phonon bath. The influence of electron-phonon coupling on tunnelling current, linear and nonlinear conductance, and on tunnel magnetoresistance is studied in detail. Variation of the main Kondo peaks and phonon satellites with the angle between magnetic moments of the leads is analyzed.

1. Introduction

Electronic transport through single molecules and quantum dots (QDs) coupled to metallic leads is a subject of current interest, particularly at low temperatures where the Kondo physics emerges [1-10]. Qualitatively new features of electronic transport appear when metallic leads are ferromagnetic and charge transport is associated with a spin current. It has been shown that the Kondo anomaly in electron transport through a quantum dot attached symmetrically to ferromagnetic leads is suppressed in the parallel configuration [11–17], while in the antiparallel configuration it has features similar to those for quantum dots attached to nonmagnetic leads. This behaviour has also been confirmed experimentally in transport measurements on large molecules [18,19] as well as on semiconductor QDs [20]. The partial or total suppression of the Kondo anomaly is a consequence of an effective exchange field originating from the spin-dependent coupling of the dot to ferromagnetic leads. It is worth noting that the suppression of Kondo anomaly may also occur in antiparallel configuration when coupling to the leads is not symmetric.

Suppression of the Kondo anomaly in electronic transport through quantum dots attached to ferromagnetic leads was studied theoretically not only for collinear magnetic configurations, but also for noncollinear ones [17]. The latter geometry is of particular interest as it gives a detailed information on the variation of the anomaly with the angle between magnetic moments of the leads. It has been shown that the Kondo peak in differential conductance is suppressed already at small deviations from the antiparallel configuration, while for larger deviations the anomaly varies rather slowly with the angle between magnetic moments.

The above mentioned theoretical and experimental investigations of the Kondo phenomenon did not take into account possible coupling of the electronic states to vibrational modes. However, recent experimental data on electronic transport through molecules and QDs reveal features which clearly indicate the role of vibrational degrees of freedom [21-29]. Extensive theoretical efforts have been undertaken to account for these observations [30-43]. However, theoretical considerations sometimes lead to contradictory predictions, particularly on the position of phonon satellites in differential conductance. It has been also reported that the nonresonant tunnelling of electrons in single-molecules is associated with excited/absorbed vibrational modes of the molecule. When the electron-phonon coupling is strong enough, tunnelling electron can absorb or emit a phonon, which has a significant influence on the transport characteristics.

In a recent paper [42] the Kondo anomaly and tunnel magnetoresistance (TMR) have been investigated in the limit of infinite Coulomb correlations on the dot coupled to ferromagnetic electrodes. It has been shown that new satellite peaks induced by the electron-phonon coupling appear on both sides of the main Kondo peak. Very recently, the same authors extended their considerations to a finite Coulomb parameter $U$ [43]. In both cases the considerations were limited to collinear magnetic configurations. In this paper we extend the theoretical considerations of the phonon-assisted electronic transport (in the Kondo regime) through quantum dots coupled to nonmagnetic leads (or ferromagnetic ones with collinear magnetizations) to the case of ferromagnetic electrodes with arbitrary orientation of the leads' magnetic moments. Since the Kondo effect in quantum dots attached to ferromagnetic leads with noncollinear alignment of the leads' moments was studied by us in Ref.[17] in the absence of electron-phonon coupling, one may also say, that the present work extends theoretical description of the Kondo effect in transport through quantum dots attached to ferromagnetic leads with noncollinear magnetizations to the case when the electrons are additionally coupled to a phonon bath.

The paper is organized as follows. In Section II we describe the system under consideration. Theoretical method is described in Section III, where the equation of motion method is used to derive nonequilibrium Green functions for the dot. Relevant numerical results for spectral functions, tunnelling current, conductance, and tunnel magnetoresistance are presented and discussed in Section IV. Summary and general conclusions are presented in Section V.

**2. Model**

The system under consideration consists of a quantum dot which is coupled to two ferromagnetic electrodes (also referred to as leads in the following). Magnetic moments of the leads are generally noncollinear and form an angle $\theta$. Electrons in the dot are coupled to a phonon bath, which includes modes of a single frequency. Hamiltonian of the system can be written in the general form as: $H = H_L + H_R + H_{ph} + H_D + H_T$, where $H_\beta = \sum_{\mathbf{k}s} \varepsilon_{\mathbf{k}\beta s} c^+_{\mathbf{k}\beta s} c_{\mathbf{k}\beta s}$ describes the left ($\beta = L$) and right ($\beta = R$) electrodes in the non-interacting quasi-particle approximation, with $c^+_{\mathbf{k}\beta s}$ ($c_{\mathbf{k}\beta s}$) being the creation (annihilation) operator of an electron in the lead $\beta$ and with the wavevector $\mathbf{k}$, energy $\varepsilon_{\mathbf{k}\beta s}$, and spin s ($s = +$ for majority electrons and $s = -$ for minority ones). The term $H_{ph} = \omega_0 a^+ a$ describes the phonon bath, with $a^+$ ($a$) being the creation (annihilation) operator of the local vibrational mode of energy $\omega_0$. Hamiltonian $H_D$ describes a single-level dot and is assumed in the form

$$H_D = \sum_\sigma E_\sigma d^+_\sigma d_\sigma + U d^+_\uparrow d_\uparrow d^+_\downarrow d_\downarrow + \lambda (a^+ + a) \sum_\sigma d^+_\sigma d_\sigma, \tag{1}$$

where $d^+_\sigma$ ($d_\sigma$) creates (annihilates) an electron with spin $\sigma$ on the dot ($\sigma = \uparrow, \downarrow$ along the quantization axis appropriate for the dot). The dot's energy level $E_\sigma$ may be spin dependent in a general case. An effective exchange field $\mathbf{B}_{ex}$ following from interaction of the dot with ferromagnetic electrodes leads to a certain renormalization and spin splitting of the dot level. Orientation of this field also determines the relevant quantization axis for the dot. In a general case, Coulomb interaction between electrons is taken into account in the Hubbard form, and is described by the parameter $U$. The last term of the Hamiltonian (1), describes the electron-phonon interaction (EPI), with $\lambda$ being the relevant coupling parameter.

The term $H_T$ of the Hamiltonian $H$ describes tunnelling processes between the leads and dot, and is assumed in the form [17]

$$H_T = \sum_{\mathbf{k}\beta} \sum_{s\sigma} W^{s\sigma}_{\mathbf{k}\beta} c^+_{\mathbf{k}\beta s} d_\sigma + h.c., \tag{2}$$

where $W^{s\sigma}_{\mathbf{k}\beta}$ are elements of the matrix $\mathbf{W}_{\mathbf{k}\beta}$,

$$\mathbf{W}_{\mathbf{k}\beta} = \begin{pmatrix} T_{\mathbf{k}\beta+} \cos(\varphi_\beta/2) & -T_{\mathbf{k}\beta+} \sin(\varphi_\beta/2) \\ T_{\mathbf{k}\beta-} \sin(\varphi_\beta/2) & T_{\mathbf{k}\beta-} \cos(\varphi_\beta/2) \end{pmatrix}, \tag{3}$$

with $\varphi_\beta$ denoting the angle between the local quantization axis in the lead $\beta$ and the quantization axis appropriate for the dot. When $\varphi_\beta = 0$, the matrix elements $T_{\mathbf{k}\beta s}$ describe electron tunnelling from the dot to the spin majority ($s = +$) and spin minority ($s = -$) electron bands in the lead $\beta$. For a symmetrical and unbiased system one may assume $\varphi_R = -\varphi_L = \theta/2$, where $\theta$ is the angle between magnetic moments of the two leads. In a general case, however, orientation of the quantization axis has to be determined self-consistently.

Coupling strength of the quantum dot and electrode $\beta$ is described by the parameter $\Gamma_s^\beta = 2\pi \sum_\mathbf{k} |T_{\mathbf{k}\beta s}|^2 \delta(E - \varepsilon_{\mathbf{k}\beta s})$. This parameter can be assumed as independent of energy within the electron band in the leads and zero otherwise. As the spin polarization of electrons at the Fermi level in the lead $\beta$ is described by a factor $p_\beta$, the coupling constants $\Gamma_s^\beta$ can be written in the form $\Gamma_s^\beta = \Gamma_0^\beta (1 + s p_\beta)$ for $s = \pm 1$.

The electric current $J$ flowing through the system is calculated according to the formula [44,45]

$$J = \frac{ie}{2\hbar} \int \frac{dE}{2\pi} Tr[(\mathbf{\Gamma}^L - \mathbf{\Gamma}^R)\mathbf{G}^<(E) + (f_L(E)\mathbf{\Gamma}^L - f_R(E)\mathbf{\Gamma}^R)(\mathbf{G}^>(E) - \mathbf{G}^<(E))], \qquad (4)$$

where $f_\beta(E)$ denotes the Fermi-Dirac distribution function for the lead $\beta$, while $\mathbf{G}^<(E)$ and $\mathbf{G}^>(E)$ are the Fourier transforms of the lesser and greater Green functions, defined as: $G^<_{\sigma\sigma'}(t) = i\langle d^+_{\sigma'}(0) d_\sigma(t) \rangle$ and $G^>_{\sigma\sigma'}(t) = -i\langle d_\sigma(t) d^+_{\sigma'}(0) \rangle$, respectively. Elements of the interaction matrix $\mathbf{\Gamma}^\beta$ can be written as follows: $\Gamma^\beta_{\uparrow\uparrow} = \Gamma^\beta_+ \cos^2(\varphi_\beta/2) + \Gamma^\beta_- \sin^2(\varphi_\beta/2)$, $\Gamma^\beta_{\downarrow\downarrow} = \Gamma^\beta_+ \sin^2(\varphi_\beta/2) + \Gamma^\beta_- \cos^2(\varphi_\beta/2)$, and $\Gamma^\beta_{\sigma\bar{\sigma}} = (1/2)(\Gamma^\beta_+ - \Gamma^\beta_-)\sin(\varphi_\beta)$.

## 3. Theoretical formulation

To determine the lesser and greater Green functions in the presence of EPI, the Hamiltonian $H$ of the system is transformed as $\tilde{H} = e^S H e^{-S} = \tilde{H}_{el} + \tilde{H}_{ph}$ with the use of the canonical transformation, $S = (\lambda/\omega_0)\sum_\sigma d_\sigma^+ d_\sigma (a^+ - a)$, which allows one to eliminate the electron-phonon coupling term from the dot Hamiltonian $H_D$ [46]. The new fermion operators are then $\tilde{d}_\sigma = d_\sigma X$ and $\tilde{d}_\sigma^+ = d_\sigma^+ X^+$, where $X = \exp[-(\lambda/\omega_0)(a^+ - a)]$. In turn, the transformed Hamiltonian $\tilde{H}$ can be written in the form $\tilde{H} = H_{ph} + H_L + H_R + \tilde{H}_D + \tilde{H}_T$, where the dot Hamiltonian has now the Anderson-type form,

$$\tilde{H}_D = \sum_\sigma \tilde{\varepsilon}_\sigma d_\sigma^+ d_\sigma + \tilde{U} d_\uparrow^+ d_\uparrow d_\downarrow^+ d_\downarrow, \qquad (5)$$

with the renormalized energy level, $\tilde{\varepsilon}_\sigma = \varepsilon_\sigma - g\omega_0$, and renormalized correlation parameter, $\tilde{U} = U - 2g\omega_0$, where $g = (\lambda/\omega_0)^2$. The tunnel Hamiltonian also becomes changed by the transformation, and the new tunnelling matrix elements are renormalized as, $T_{k\beta s} \Rightarrow \tilde{T}_{k\beta s} = T_{k\beta s} X$.

The electron and phonon subsystems become decoupled, when the phonon operator $X$ appearing in $\tilde{H}_T$ is replaced by its expectation value in thermal equilibrium, $X \Rightarrow \langle X \rangle = \exp[-g(N_{ph} + 1/2)]$, where $N_{ph}$ denotes the equilibrium phonon population. The transformed Hamiltonian, $\tilde{H}$, can be subsequently used to determine time evolution of the system.

The relevant lesser Green function can be expressed in the following form:

$$G_{\sigma\sigma'}^<(t) = i\langle d_{\sigma'}^+(0) d_\sigma(t)\rangle_H = i\langle d_{\sigma'}^+(0) X^+(0) d_\sigma(t) X(t)\rangle_{\tilde{H}}, \qquad (6)$$

where the subscripts $H$ and $\tilde{H}$ indicate the appropriate Hamiltonian that governs the system evolution. Since the electron and phonon subsystems are decoupled, the corresponding average values can be calculated independently. Accordingly, the Green function $G_{\sigma\sigma'}^<(t)$ can be expressed as [38]

$$G^{<}_{\sigma\sigma'}(t) = i\langle d^{+}_{\sigma'}e^{i\tilde{H}_{el}t}d_{\sigma}e^{-i\tilde{H}_{el}t}\rangle\langle X^{+}e^{iH_{ph}t}Xe^{-iH_{ph}t}\rangle = \tilde{G}^{<}_{\sigma\sigma'}e^{-\Phi(-t)}, \tag{7}$$

where $\Phi(t) = g[N_{ph}(1-e^{i\omega_0 t}) + (N_{ph}+1)(1-e^{-i\omega_0 t})]$. The Fourier transform of the lesser Green function is then equal

$$G^{<}_{\sigma\sigma'}(E) = \sum_{n=-\infty}^{\infty} L_n \tilde{G}^{<}_{\sigma\sigma'}(E + n\omega_0) \tag{8}$$

with $L_n = e^{-g(2N_{ph}+1)}e^{n\omega_0/2k_BT}I_n(2g\sqrt{N_{ph}(N_{ph}+1)})$, where $I_n(z)$ is the n-th Bessel function of complex argument. Similarly, writing $G^{>}_{\sigma\sigma'}(E)$ in the form

$$G^{>}_{\sigma\sigma'}(E) = \sum_{n=-\infty}^{\infty} L_n \tilde{G}^{>}_{\sigma\sigma'}(E - n\omega_0), \tag{9}$$

one can calculate the spectral function as

$$A_{\sigma}(E) = i(G^{>}_{\sigma\sigma} - G^{<}_{\sigma\sigma}). \tag{10}$$

We point, that $\tilde{G}^{<(>)}_{\sigma\sigma'}$ is determined with the use of Hamiltonian $\tilde{H}_{el}$ which has the form similar to that describing a single-level dot attached to external electrodes via tunnel terms. However, the key parameters of the model are renormalized due to the presence of phonons. Thus, the general relations derived for the Green functions in Ref.[17] also hold in this particular case, so the Green functions $\tilde{G}^{<(>)}_{\sigma\sigma'}$ can be easily found. In particular, the lesser Green function can be determined from the Keldysh equation, $\tilde{\mathbf{G}}^{<} = \tilde{\mathbf{G}}^{r}\mathbf{\Sigma}^{<}\tilde{\mathbf{G}}^{a}$, with the retarded $\tilde{\mathbf{G}}^{r}$ and advanced $\tilde{\mathbf{G}}^{a}$ Green functions calculated from the equation of motion within the decoupling scheme appropriate for the Kondo regime.

To describe spin splitting of the dot level due to ferromagnetism of the external electrodes we introduce an effective exchange field $\mathbf{B}_{ex}$ which is exerted on the dot by the electrodes. Such an approach leads to results which are in agreement with those obtained by other ways of introducing spin splitting of the dot levels [11,15]. The exchange field $\mathbf{B}_{ex}$ is calculated according to the formula [17]:

$$\mathbf{B}_{ex} = \frac{1}{g\mu_B}\sum_\beta \mathbf{n}_\beta \operatorname{Re}\int \frac{d\varepsilon}{2\pi}(\tilde{\Gamma}_+^\beta - \tilde{\Gamma}_-^\beta)\frac{f_\beta(\varepsilon)}{\varepsilon - \tilde{\varepsilon}_0 - i\hbar/\tau_0}\,, \tag{11}$$

where $\mathbf{n}_\beta$ is the unit vector along magnetic moment of the electrode $\beta$, $\tau_0$ is the relevant relaxation time, whereas $\tilde{\varepsilon}_0$ and $\tilde{\Gamma}_s^\beta$ stand for the spin-degenerate energy level and coupling strength to the lead $\beta$, respectively, which are renormalized by the electron-phonon coupling. Strictly speaking, $\tilde{\Gamma}_s^\beta$ is defined in a similar way as $\Gamma_s^\beta$, but with $\tilde{T}_{k\beta s}$ instead of $T_{k\beta s}$. Thus, the coupling is reduced when the vibrational modes are taken into account.

The self-energy $\tilde{\Sigma}^<$ which enters the Keldysh equation is calculated from the Ng ansatz as: $\tilde{\Sigma}^< = \tilde{\Sigma}_0^<(\tilde{\Sigma}_0^r - \tilde{\Sigma}_0^a)^{-1}(\tilde{\Sigma}^r - \tilde{\Sigma}^a)$, where $\tilde{\Sigma}_0^< = i(\tilde{\Gamma}^L f_L + \tilde{\Gamma}^R f_R)$ is the lesser self-energy of the corresponding non-interacting system with the matrix elements $\tilde{\Gamma}_{\sigma\sigma'}^\beta$ defined in a similar way as $\Gamma_{\sigma\sigma'}^\beta$, and $\tilde{\Sigma}_0^r - \tilde{\Sigma}_0^a = -i\tilde{\Gamma}^\beta = -i(\tilde{\Gamma}^L + \tilde{\Gamma}^R)$. The retarded and advanced self-energies of the correlated system can be calculated from the Dyson equation $(\mathbf{I} - \tilde{\mathbf{g}}_0\tilde{\Sigma})\tilde{\mathbf{G}} = \tilde{\mathbf{g}}_0$, and may be written in the form $\tilde{\Sigma} = \tilde{\mathbf{g}}_0^{-1} - \mathbf{n}^{-1}\tilde{\mathbf{g}}_0^{-1} + \mathbf{n}^{-1}\tilde{\Sigma}_w$, where $\tilde{g}_{0\sigma\sigma'} = \delta_{\sigma\sigma'}(E - \tilde{\varepsilon}_0)^{-1}$, $n_{\sigma\sigma} = 1 - <d_{\bar{\sigma}}^+ d_{\bar{\sigma}}>$, $n_{\sigma\bar{\sigma}} = -<d_{\bar{\sigma}}^+ d_\sigma>$, with $\bar{\sigma} = -\sigma$ and $\tilde{\Sigma}_w = \tilde{\Sigma}_0 + \tilde{\Sigma}_1$. All processes which lead to the Kondo effect are included into $\tilde{\Sigma}_1$, which for $U \Rightarrow \infty$ is defined as follows: $\tilde{\Sigma}_{1\sigma\sigma} = \sum_\beta \int \frac{d\varepsilon}{2\pi}\frac{\tilde{\Gamma}_{\bar{\sigma}\bar{\sigma}}^\beta f_\beta(\varepsilon)}{E - \tilde{\varepsilon}_\sigma + \tilde{\varepsilon}_{\bar{\sigma}} - \varepsilon + i\hbar/\tau_{\bar{\sigma}}}$ and $\tilde{\Sigma}_{1\sigma\bar{\sigma}} = \sum_\beta \int \frac{d\varepsilon}{2\pi}\frac{\tilde{\Gamma}_{\bar{\sigma}\sigma}^\beta f_\beta(\varepsilon)}{E - \varepsilon}$. The procedure briefly outlined above allows us to calculate the Green functions $\tilde{G}_{\sigma\sigma'}^<(E)$ and $\tilde{G}_{\sigma\sigma'}^>(E)$. Then, the Green functions $\mathbf{G}^<$ and $\mathbf{G}^>$ can be determined according to Eqs (8) and (9). We note that the Green functions and the relevant occupation numbers, $<d_\sigma^+ d_\sigma> = -i\int \frac{dE}{2\pi}G_{\sigma\sigma}^<$, have been calculated self-consistently. Finally, having found the Green functions, one can calculate current flowing through the system using Eq.(4).

## 4. Numerical results

We consider now some numerical results obtained within the formalism outlined above. For convenience, we will use the relative energy units, and as the unit we assume D/50, where D is the electron band width. For numerical calculations we assume the following values of the parameters: the bare dot level $E_\sigma = E_0 = -0.31$, phonon energy $\omega_0 = 0.05$, leads polarization p=0.2, and coupling parameters $\Gamma_L = \Gamma_R = 0.1$. We will consider two different situations as concerns strength of the electron-phonon coupling – weak coupling corresponding to $g = 0.1$, and strong coupling corresponding to $g = 0.4$.

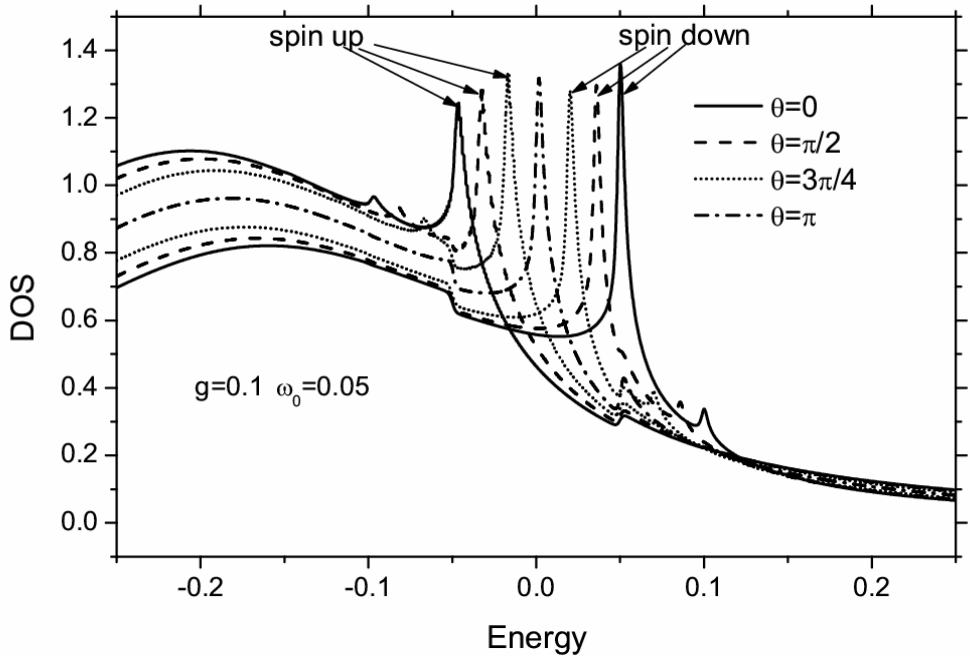

Figure1. DOS in equilibrium situation for both spin orientations and for indicated values of the angle between magnetic moments. The other parameters are: $E_0 = -0.31$, $\omega_0 = 0.05$, $\Gamma_L = \Gamma_R = 0.1$, p=0.2, $k_B T = 0.001$, and $g = 0.1$. For $\theta = \pi$ the peaks for both spin orientations overlap.

In figure 1 we show the equilibrium (zero bias) density of states (DOS) for both spin orientations (along the quantization axis of the dot) in the case of weak electron-phonon coupling and for indicated values of the angle between magnetic moments. Since the dot is symmetrically coupled to the leads, the Kondo peak in DOS in the antiparallel configuration ($\theta = \pi$) appears at the Fermi level independently of the spin orientation. This is a consequence of the compensation of effective exchange fields from the two leads in this particular magnetic configuration. Apart from this, the phonon satellites in DOS appear on both sides of the main peak at the distance equal to $\omega_0$. Such a behaviour is similar to that obtained for a system with non-magnetic electrodes [41].

When the configuration departs from the antiparallel one, the main Kondo peaks (i.e. those in the absence of electron-phonon coupling) become shifted away from the Fermi level; to the right (higher energy) for spin-down orientation and to the left (lower energy) for the spin-up orientation. The phonon satellites in DOS for spin-up (spin-down) orientation appear on the left (right) side of the corresponding main resonance and move together with this resonance when the angle varies from antiparallel to parallel orientation. The splitting of the main Kondo peak in DOS for noncollinear configurations is a consequence of a nonzero effective exchange field exerted by the leads on the dot in such configurations [17]. The situation is very similar to that for a non-magnetic system in an external magnetic field, when a finite Zeeman splitting of the dot level leads to splitting of the Kondo peak. Each phonon satellite peak moves coherently with the corresponding component of the main peak. Apart from this, the splitting of the Kondo anomaly monotonically increases as the angle $\theta$ between magnetic moments of the leads varies from $\theta = \pi$ (antiparallel configuration) to $\theta = 0$ (parallel configuration).

For the assumed values of the parameters (particularly of the electron-phonon coupling strength) the shifts of the peaks corresponding to the up and down spin orientations are of the order of phonon energy $\omega_0$ in the parallel configuration. Accordingly, the main Kondo resonances in this configuration appear roughly for energies $\pm \omega_0$, whereas the corresponding satellite peaks develop near $\pm 2\omega_0$ with respect to the Fermi level. Since the electron-phonon coupling in figure 1 is relatively weak, the main Kondo resonances are well pronounced.

Apart from the phonon satellite Kondo peaks, coupling of the vibrational modes to electrons also slightly modifies the spectral function. More specifically, the electron-phonon coupling also leads to jumps in DOS at the energies $\pm n\omega_0$, independently of the magnetic configuration of the system. These steps result from the Fermi distribution function and are very sharp at low temperatures. However, they become washed out when the temperature increases. In figure 1 these steps are clearly seen for *n*=1, while those corresponding to higher *n* are not resolved due to weak electron-phonon coupling.

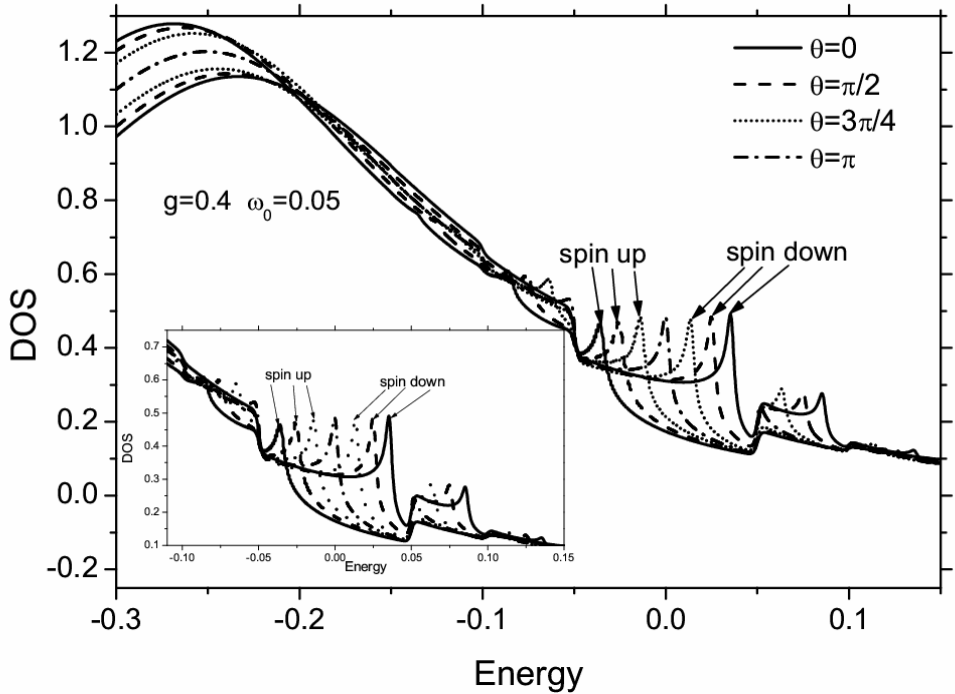

Figure 2. DOS in equilibrium situation for both spin orientations and for indicated values of the angle between magnetic moments. The other parameters are: $E_0 = -0.31$, $\omega_0 = 0.05$, $\Gamma_L = \Gamma_R = 0.1$, p = 0.2, $k_BT = 0.001$, and g = 0.4. The inset shows the part of the spectrum around the Kondo peaks.

A different situation is presented in figure 2, where DOS is shown for the case of strong electron-phonon coupling, corresponding to g = 0.4, while all the other parameters are the same as in figure 1. The spectral function is now dominated by the broad maximum which appears at the renormalized dot's energy level $\tilde{\varepsilon}_\sigma$. The main Kondo resonances are clearly visible, although their intensities are considerably reduced due to the electron-phonon coupling. At the same time, the satellite peaks are well pronounced, as can be seen in the inset to figure 2, where only the part of spectrum around the Kondo peaks is shown. Now, not only the first phonon satellite Kondo peaks are resolved, but also the second ones, separated by $2\omega_0$ from the main Kondo peaks, are clearly seen.

By comparison of figures 1 and 2 one can note that position of the main Kondo peaks is slightly modified by the electron-phonon coupling. More precisely, the splitting of the Kondo peak in figure 2 is slightly reduced in comparison to that in figure 1. We note, that this splitting is determined by the exchange field exerted by the leads on the dot. According to Eq. (10), this exchange field depends on the dot-lead coupling parameters, which are renormalized by the electron-phonon interaction. Since the effective dot-lead coupling strengths are considerably reduced by the electron-phonon coupling, the splitting of the main Kondo resonance in figure 2 is smaller than that in figure 1, in agreement with Ref. [43]. In turn, the corresponding phonon satellite peaks are better resolved.

The Kondo peaks in DOS become additionally split in nonequilibrium situations, when a bias voltage is applied. This is well known also in other situations, so we will not discuss the problem here in more details. Instead of this we will consider now electron transport in a biased system, where the Kondo peaks in DOS lead to anomalous behaviour of the differential conductance in the small voltage regime (zero bias anomaly).

In figure 3 we show differential conductance (only the total conductance is shown there) in the case of weak electron-phonon coupling and for indicated values of the angle between magnetic moments. In the antiparallel configuration the Kondo peak appears at the zero bias limit, $V = 0$. Moreover, two phonon peaks develop on both sides of the main peak at $\pm\omega_0$. When the magnetic configuration departs from the antiparallel one, the main Kondo peak becomes split and the two components move away from the zero bias limit (one towards positive and the other towards negative bias). The two additional and well-resolved phonon peaks remain at the energies $\pm\omega_0$, independently of magnetic configuration. These peaks follow from the steps in DOS at $\pm\omega_0$, and can be interpreted as the onset of phonon assisted channel for tunnelling. Typical phonon Kondo peaks appear at the distance $\omega_0$ from the main Kondo peaks. However, these features are not well resolved in figure 3 due to weak electron-phonon coupling. Moreover, for the parameters assumed in figure 3, the Kondo peaks in the parallel configuration overlap with the peaks at $\pm\omega_0$, so only a single peak of enhanced intensity is observed for each voltage polarization in this configuration.

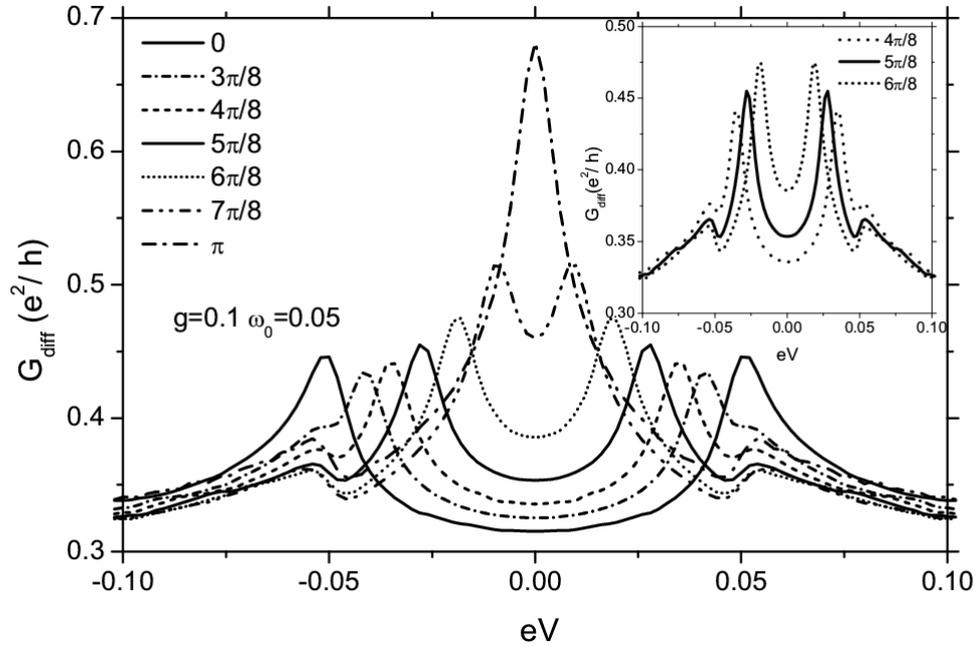

Figure 3. Differential conductance for indicated values of the angle between magnetic moments. The other parameters are: $E_0 = -0.31$, $\omega_0 = 0.05$, $\Gamma_L = \Gamma_R = 0.1$, $p = 0.2$, $k_BT = 0.001$, and $g = 0.1$. Some of the curves are shown in the inset for a broader voltage range, which allows identification of the Kondo peaks.

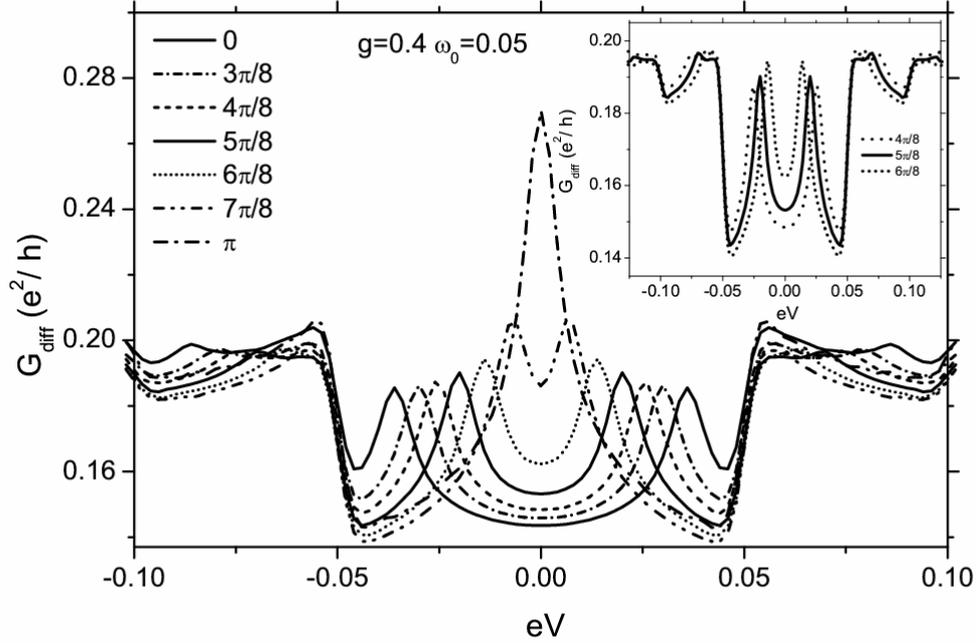

Figure 4. Differential conductance for indicated values of the angle between magnetic moments. The other parameters are: $E_0 = -0.31$, $\omega_0 = 0.05$, $\Gamma_L = \Gamma_R = 0.1$, $p = 0.2$, $k_BT = 0.001$, and $g = 0.4$. Some of the curves are presented in the inset for a broader voltage range.

In figure 4 we show differential conductance in the case of strong electron-phonon coupling. The key qualitative difference is that now the effective exchange field exerted on the dot is reduced by the electron-phonon coupling, and accordingly the Kondo peaks in the parallel configuration do not overlap with the peaks at $\pm\omega_0$, and both are clearly seen in the conductance. Intensities of the main Kondo peaks, however, are considerably reduced in comparison to those in the case of weak electron-phonon coupling, and the low-bias Kondo anomaly is suppressed. The suppression of the peaks takes place for all magnetic configurations. Moreover, apart from the main components, the two phonon satellite Kondo peaks which develop in a distance $\omega_0$ from each anomaly are now well resolved. Positions of these satellites depend on the angle $\theta$ and they move coherently with the main peaks as $\theta$ is changed. This behaviour can be clearly seen in the inset to figure 4, where the conductance is depicted in a broader energy region and only for a few values of $\theta$.

Difference in transport characteristics for parallel and noncollinear geometries gives rise to the TMR effect, which can be described quantitatively by the ratio TMR $(\theta)$ = $[I(\theta = 0) - I(\theta)]/I(\theta)$, where $I(\theta)$ denotes the current flowing through the system when magnetic moments of the leads form an angle $\theta$. The numerical results for TMR, corresponding to figures 3 and 4, are shown in figure 5.

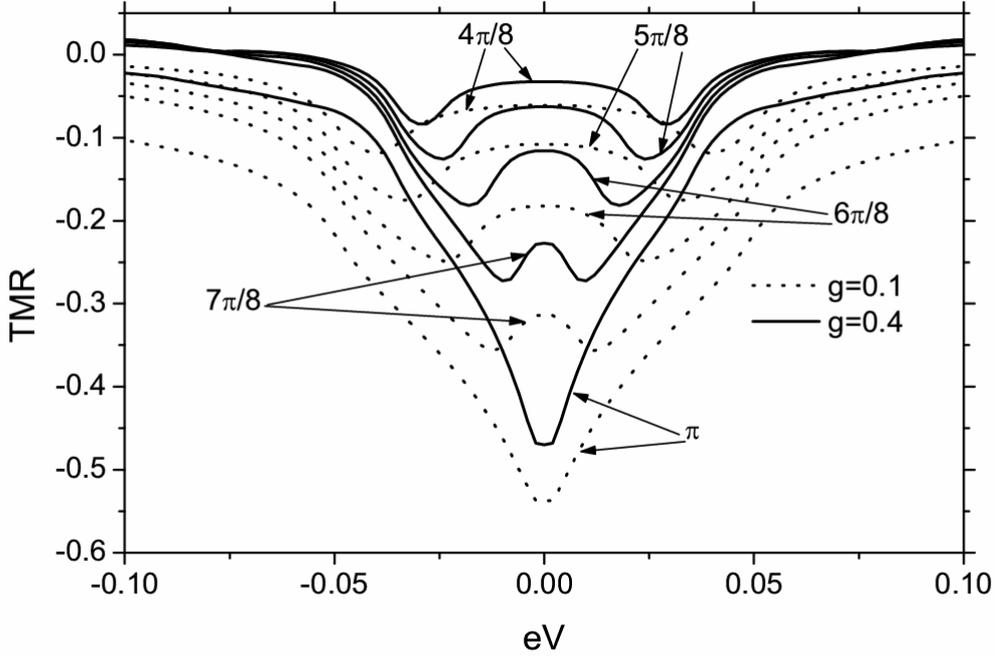

Figure 5. TMR effect for weak and strong electron-phonon coupling and for parameters as in figure 2 and 3, respectively.

The main feature of TMR is that it becomes negative at small voltages. Similar negative TMR was also found in systems without electron-phonon coupling [17], so we will not discuss this point more as its origin is the same. Instead, we focus on the effects due to electron-phonon interaction. We point, that the shape of the TMR curve is determined mainly by the Kondo anomaly and resembles the one obtained in the absence of EPI. Although, the electron-phonon interaction leads to well resolved satellites in the differential conductance, the features of electron-phonon coupling are only weakly seen in the shape of the TMR curve. However, the electron-phonon interaction has a significant influence on the TMR magnitude. Generally, for strong electron-phonon coupling, the TMR magnitude decreases with increasing coupling and becomes close to zero in a wide range of bias voltages. This is shown explicitly in figure 6, where the angular dependence of TMR is shown for several values of bias voltage applied to the system, and for two values of the electron-phonon coupling strengths.

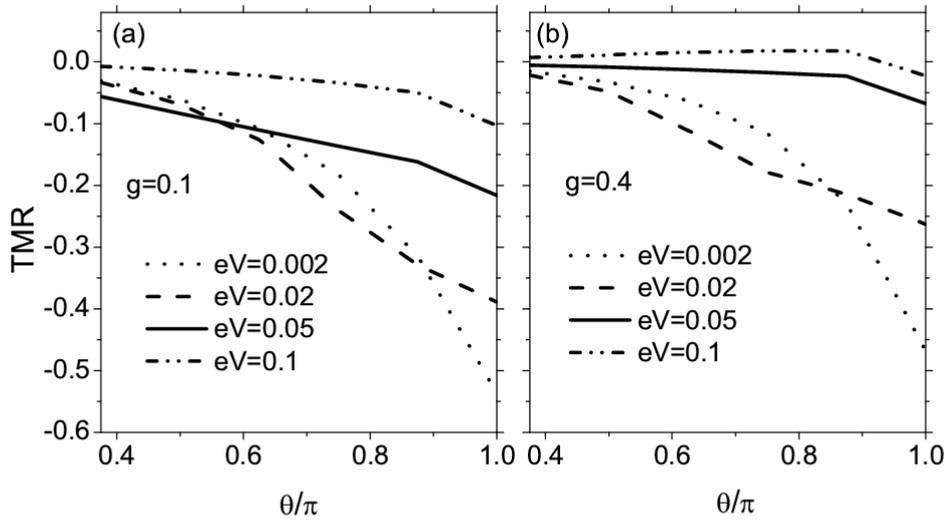

Figure 6. TMR versus the angle $\theta$ for indicated values of the bias voltage, and for two different values of the coupling parameter g=0.1 (a) and g=0.4 (b).

From figures 5 and 6 follows that considerable changes of TMR occur mainly in the regime of small voltages. Moreover, absolute magnitude of TMR increases then in a monotonous way with the angle $\theta$ between magnetic moments. The increase in TMR is particularly fast when the system approaches the antiparallel configuration. For higher voltages, on the other hand, TMR is rather small. Moreover, it changes sign from negative (inverse spin valve) to positive (normal spin valve). This behaviour is similar to that found in systems without electron-phonon interaction [17].

## 5. Summary and concluding remarks

We have studied the influence of electron-phonon interaction on spin polarized transport through a single-level quantum dot in the Kondo regime. The interplay of the effects resulting from spin-dependent coupling of the dot to external ferromagnetic electrodes and due to electron-phonon interaction in the dot has been analyzed in detail. The main features due to the vibrational modes are the phonon satellite peaks which develop in the density of states. These additional peaks accompany the main spin-dependent Kondo components and move accordingly when magnetic configuration of the system varies continuously from antiparallel to parallel alignment. The Kondo satellites due to electron-phonon coupling are not very well pronounced in the differential conductance and can be observed only in the case of strong electron-phonon interaction. The phonon Kondo satellites move then coherently with the main resonances as magnetic configuration is changed. Electron-phonon coupling has only a weak influence on the shape of the TMR curve as a function of the bias voltage. However, the TMR effect becomes significantly reduced by a strong electron-phonon interaction.

**Acknowledgements**

The authors thank W. Rudzinski for useful discussions. This work was supported by funds of the Polish Ministry of Science and Higher Education as a research project in years 2006-2009.